\documentclass[runningheads]{llncs}
\usepackage{tabularx}
\usepackage{booktabs}
\usepackage[most]{tcolorbox}
\usepackage{amssymb}
\usepackage{enumitem}
\usepackage{etoolbox}
\usepackage[T1]{fontenc}
\usepackage{graphicx}
\tcbset{
  flowbox/.style={
    colback=#1!15!white,
    colframe=#1!70!black,
    fontupper=\small,
    halign=center,
    valign=center,
    arc=3pt,
    width=0.85\linewidth,
    boxrule=1pt,
    top=6pt, bottom=6pt
  }
}
\begin{document}
\title{Digital White Spaces: A Cyberpsychology-Informed Framework to Mobile Phone Addiction}
%
%
\author{Leandros Maglaras\inst{1,}\inst{2}\and
Chrstina Kyritsi\inst{2}, Helge Janicke\inst{3}\and Konstantinos Karantzalos\inst{2,}\inst{4}}

\authorrunning{Maglaras L. et al.}

\institute{School of Computing, Edinburgh Napier University, Edinburgh UK \and
Ministry of Digital Governance, Greece \and
School of Computer Science,  Edith Cowan University, Perth, Australia \and Remote Sensing Lab, National Technical University of Athens, Athens, Greece}
\maketitle              
\begin{abstract}
Mobile-phone overuse and attention fragmentation have become pressing societal and public-health concerns. Cyberpsychology research highlights addictive engagement loops driven by intermittent rewards, persuasive design, and habit formation. In this article, we use current evidence on mobile-phone addiction and propose "Digital White Spaces" (DWS), a socio-technical framework that combines privacy-preserving monitoring, AI-driven detection of addictive loops, device-mode interventions, and physical signal-limited zones to minimize  digital stimulation and
internet addiction.

\keywords{Cyber Addiction; Cyberpsychology \and Digital White Spaces \and Business Email Compromise (BEC)}
\end{abstract}
\section{Introduction}
\label{sec-intro}

The first GSM (Global System for Mobile Communications) network was launched in Finland in 1991, a region characterized by sparse populations and vast, isolated landscapes where frequent face-to-face communication was physically difficult. By adopting a digital wireless standard, these communities could bypass the need for physical cables, providing a reliable lifeline for safety and business in areas where isolation was a reality. Moreover, the GSM  network was heavily driven by Scandinavian countries since they were already world leaders in mobile technology, having pioneered the an automatic, international cellular network, the Nordic Mobile Telephony (NMT).  Although GSM became a pan-European standard (and eventually global), the initiative was heavily championed by the Scandinavian countries, managed to lead in the development of the new standard \cite{manninen2002elaboration}.  Meanwhile, mobile phones evolved from the 1970s analog prototypes to modern, AI-enabled 5G smartphones. Key milestones include the first call in 1973, commercialization in 1983, the introduction of SMS/2G in the 1990s, and the smartphone revolution launched by iPhone in 2007. They rapidly transitioned from business tools to global communication devices \cite{farley2005mobile}. 

The everyday use of smartphones has transformed communication and productivity, but also created conditions for uncontrollable use, attention fragmentation, and affected well-being \cite{cemiloglu2022combatting}.  Following the term of the {\bf online self} as described in \cite{attrill2019online}, every person who is using the internet demonstrates an online behavior that determines this.  Every action that a person takes during any observation or interaction with others inside the online digital world, outlines the online self. Sometimes. the internet makes it easy for people to fake a perfect life, creating a fake person, just to show off, grow the number of their followers, and cash in on this extra visibility

Cyberpsychology studies show that design features (notifications, infinite scroll, variable rewards) exploit neural reinforcement mechanisms, producing frequent task-switching and elevated stress and anxiety \cite{swikatek2023problematic}. Clinical and behavioral measures increasingly characterize problematic smartphone use as an emergent public-health challenge requiring multidisciplinary responses that combine behavioral science, engineering, and policy. 

As stated in \cite{mostyn2025exploration}, problematic smartphone use (PSU), defined as excessive or uncontrolled smartphone behavior, may lead to harm or limited daily functioning. {\bf While mobile phones are advertised as tools for constant connectivity on the move, they have recently become digital traps, anchoring users to a virtual universe while they remain motionless in the physical world.} There are everyday examples of people of all ages using their mobile phones for several hours, standing still on their couches, and thus selling their physical life to the digital fake paradise. 

Recent initiatives, such as Greece's national strategy, emphasize age verification, reducing algorithmic manipulation, and encouraging healthy digital habits. Similarly, the Ministry of Health in Singapore issued in 2025 updated guidelines regarding screen use for children aged until 12 years \cite{signapure}. Protecting minors from internet addiction involves a combined approach of setting strict screen time limits, fostering open communication, using parental control tools, and promoting alternative offline activities \cite{toska2024computer}.  But minors are not the only problem. People of all ages are lured into this fake happiness, and drastic actions needed to be taken right away.

A promising response is to move beyond individual “self-control” narratives and toward structural interventions that reduce the use of digital devices and online environments \cite{setia2025digital}. The Digital White Spaces (DWS) concept proposed in this article offers a socio-technical approach by pairing ethical, privacy-preserving sensing and AI assistance,  with device-level reduction modes and spatial interventions that can suppress digital stimulation and internet addiction.

\section{Definitions and Background}

In order to better understand the current situation that has arisen from the use of smartphones and the cyberscychology aspects of it we briefly present some core definitions used in this paper.
\begin{itemize}
    
\item {\bf Problematic Smartphone Use (PSU)}: An umbrella construct defined as repetitive, excessive, or non-regulated smartphone interaction that affects daily functioning, sleep quality, and psychological well-being, without necessarily meeting the full psychiatric criteria for clinical behavioral addiction \cite{pivetta2019problematic}.

\item {\bf Mobile Phone Addiction}: A severe, pathological form of behavioral technology dependence characterized by diagnostic addiction criteria—specifically compulsion, loss of control, salience, tolerance, withdrawal distress when restricted from the device, and continued overuse despite major personal or functional impairments \cite{kwon2013smartphone}.

\item {\bf Ghost Mode}: A software or device-level structural intervention layer that systematically removes persuasive design features (such as push notifications, infinite feeds, and data-intensive applications) while restricting the device strictly to essential communications (voice calls and SMS) and utility functions to facilitate voluntary digital detox \cite{radtke2022digital}, \cite{setia2025digital}.

\item {\bf Points of No Return}: Behavioral transition thresholds in cyberpsychology where repetitive checking habits transition into automatic, cue-triggered execution; at these points, cognitive self-control and goal-directed monitoring drop significantly, making disengagement without external environmental friction or structural barriers extremely difficult \cite{wood2022habits}).
\end{itemize}

Citizens face persistent information overload, cognitive fatigue, notification dependency, attention fragmentation, and diminished face-to-face interpersonal connection. While recent governmental frameworks have focused on child protection (such as national strategies for youth online safety or digital wallet restrictions), there remains a critical policy gap surrounding adult digital wellbeing and healthy technology management. 

Just as modern smart cities provide urban parks for physical health, they must offer designated sanctuaries for cognitive and psychological restoration. Digital White Spaces can operate under three core principles:
\begin{itemize}
    \item Voluntariness: Participation must be entirely user-led
    \item Choice: Users choose their level and duration of disconnection.
    \item Inclusivity: Zones serve as "Digital Wellbeing Zones" rather than anti-technology "No Internet" exclusions
\end{itemize}

\subsection {Theoretical Framework: The COM-B Model for Behavior Change}
Sustainable behavioral modification requires aligning capability, opportunity, and motivation. The proposed DWS explicitly map onto Michie et al.’s COM-B Model of Behavior Change \cite{michie2011behaviour}:

\begin{itemize}
    
\item Capability (Psychological \& Physical): Digital overload diminishes executive self-control. Ghost Mode Zones build user self-regulation skills, mindfulness, and cognitive capacity by providing clear guidelines, digital literacy prompts, and structured offline activities.

\item Opportunity (Physical \& Social): The modern urban environment offers ubiquitous connectivity without physical barriers to digital noise. DWS provide the physical infrastructure (parks, libraries, quiet corridors) and social context (peer participation, normalized offline behavior) necessary to make disconnection feasible.

\item Motivation (Reflective \& Automatic): By linking disconnection with tangible intrinsic rewards—such as reduced anxiety, deeper focus, and enhanced interpersonal connection—citizens develop positive attitudes and habitual drive toward mindful technology use.
\end{itemize}

Beyond COM-B, DWS use Rogers’ Diffusion of Innovations  (serving as safe trial environments for adopting digital detox habits) and marketing communication models like the Hierarchy of Effects (moving users from Awareness to Preference and Action - See Figure \ref{fig:awareness_hierarchy}).

\begin{figure}[!htbp]
    \centering
    \includegraphics[width=0.9\linewidth]{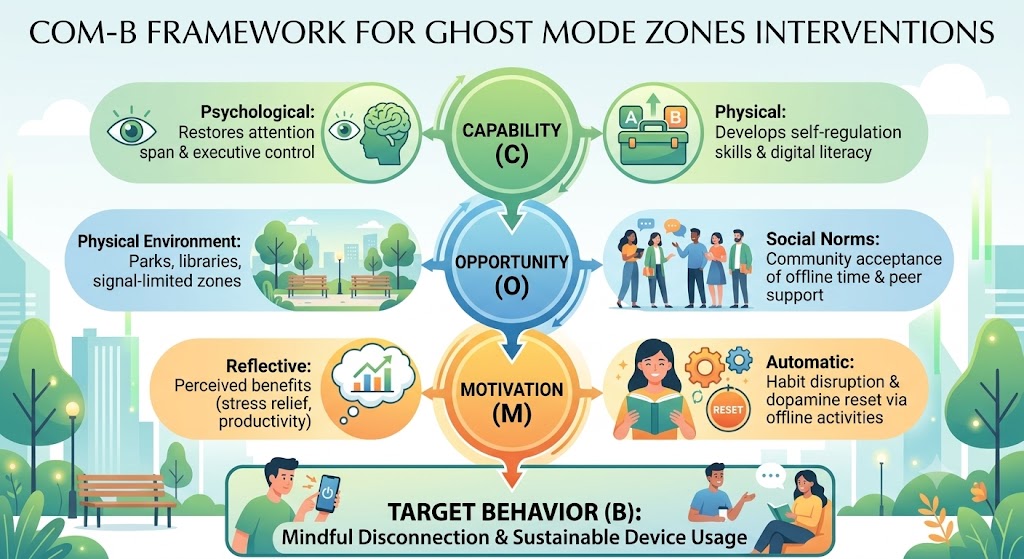}
    \caption{COM-B Framework}
    \label{fig:COMB}
\end{figure}

\begin{figure}[htbp]
\centering
\begin{tcolorbox}[flowbox=blue, colback=blue!30!white]
  \textbf{1. Awareness}\\
  Building basic recognition of DWS and digital wellbeing practices
\end{tcolorbox}

\vspace{-0.2cm}
$\downarrow$
\vspace{-0.2cm}

\begin{tcolorbox}[flowbox=blue, colback=blue!20!white]
  \textbf{2. Knowledge}\\
  Understanding the cognitive \& psychological benefits
\end{tcolorbox}

\vspace{-0.2cm}
$\downarrow$
\vspace{-0.2cm}

\begin{tcolorbox}[flowbox=cyan]
  \textbf{3--5. Affective Shift (Liking, Preference, Conviction)}\\
  Developing positive attitudes toward physical disconnection
\end{tcolorbox}

\vspace{-0.2cm}
$\downarrow$
\vspace{-0.2cm}

\begin{tcolorbox}[flowbox=green]
  \textbf{6. Action / Adoption}\\
  Active participation \& habit formation in DWS
\end{tcolorbox}

\caption{The Awareness stage mapped within the Hierarchy of Effects Model for DWSs.}
\label{fig:awareness_hierarchy}
\end{figure}

\section{Digital White Spaces: Concept and Components}
\label{sec-manipulation}	 

Digital White Spaces (DWS) are deliberate, bounded reductions in non-essential digital connectivity and persuasive measures designed to help people gain back their physical life. DWS combine both digital measures with physical boundaries. On the latter, a DWS can be either a non-connected separate place in a house, a dedicated time digital free period per day, signal-limited  public libraries or other similar places, or finally, whole city areas or walking paths that are free of any internet connection. 

\begin{figure}[!htbp]
    \centering
    \includegraphics[width=0.9\linewidth]{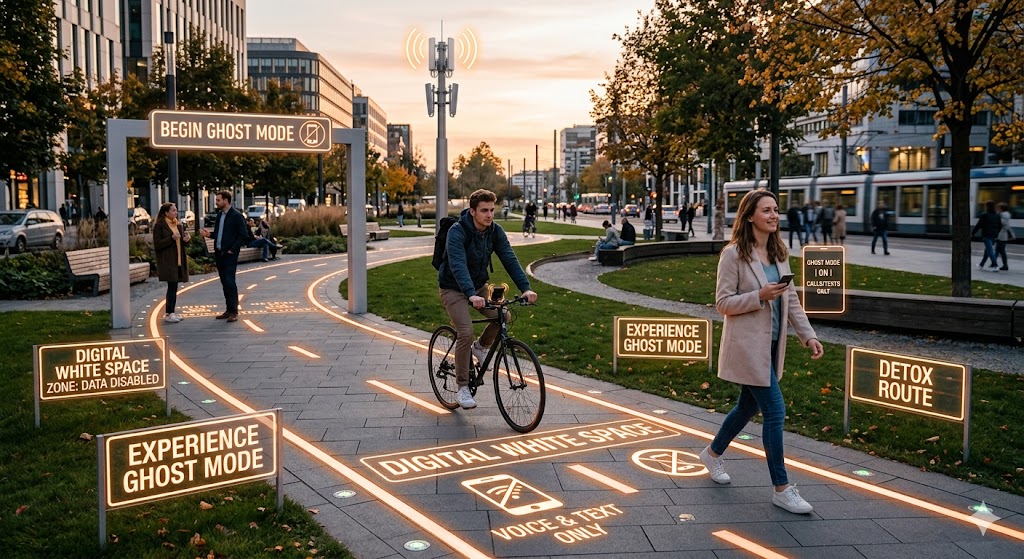}
    \caption{DWS conceptual image}
    \label{fig:DWC}
\end{figure}

\begin{figure}[!htbp]
    \centering
    \includegraphics[width=0.9\linewidth]{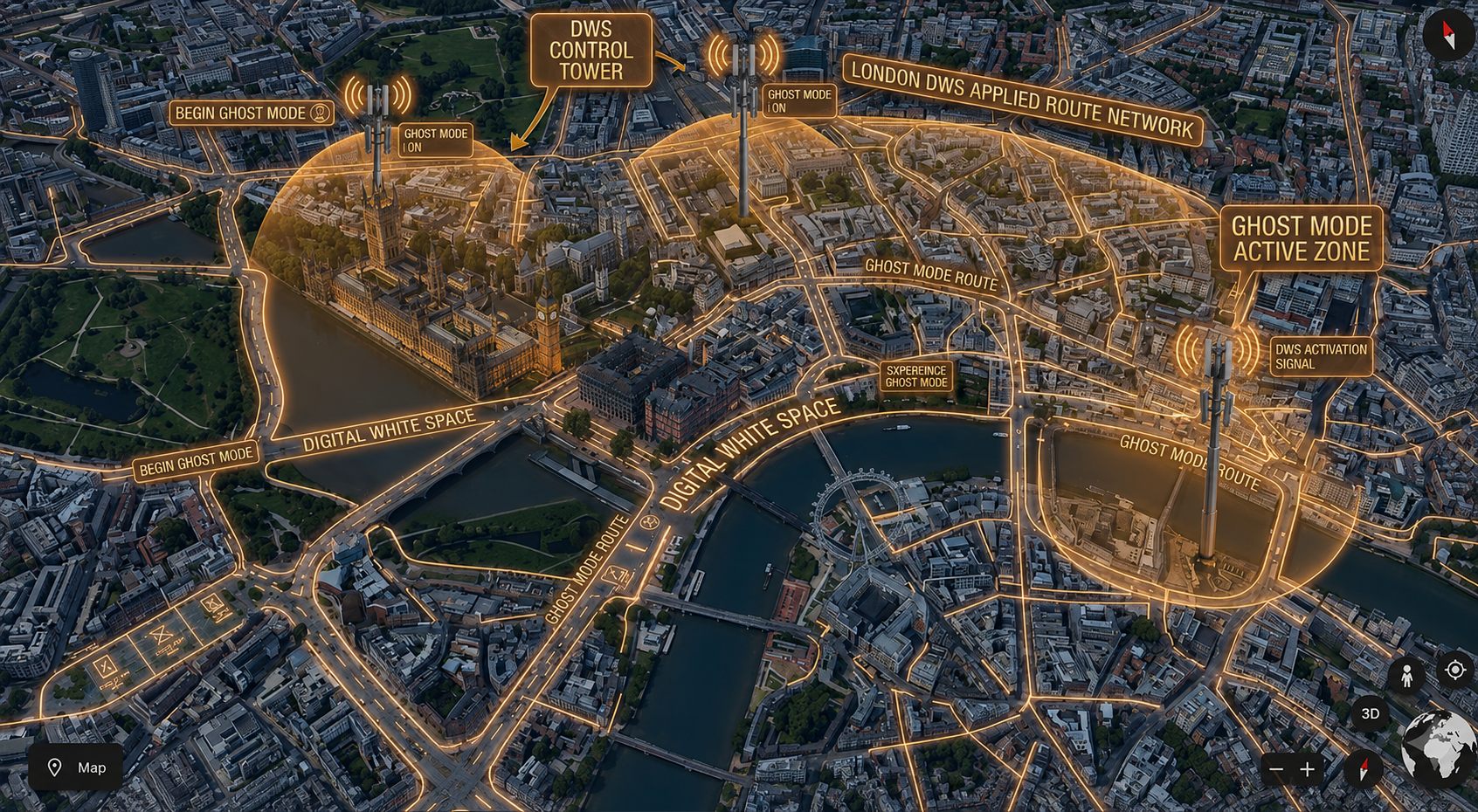}
    \caption{DWS concept over the center of London City}
    \label{fig_GPS}
\end{figure}

In Figure \ref{fig:DWC}, a conceptual image of the DWS route in a smart city is shown. In this image, the "Digital White Space" is represented as an architectural route integrated into a smart city. Along this pathway, glowing boundaries define the zone where all modern data services are deactivated, restricting mobile devices to only basic calls and text messages while holographic icons of blocked apps highlight it. As shown in Figure \ref{fig_GPS}, DWSs can be implemented at a city-wide scale, ranging from individual routes to entire districts thus allowing participating users to experience designated periods of internet disconnection.

\subsection{Core components}

On the core of the DWS idea lies a combination of sensors, tailored AI assistants, dedicated mobile phone applications, and specific telecommunication solutions. 
The main components that can be used alone or combined are:

\begin{itemize}

\item {\bf Privacy-oriented human sensors and edge AI}. On-device monitoring of usage patterns (session durations, app switches, scroll dynamics) that extracts behavioral features without revealing the exact content to others. Private information must be kept at a local level, and only aggregated information can be uploaded to the cloud. Local AI models will use this data in order to detect habits toward “points of no return” in addictive loops.

\item {\bf Data mining and AI decision layer}. All data collected from the individuals, combined with their profiles and current needs, can be used to propose the best interventions and counteractions.

\item {\bf Adaptive device interventions}. These can be “Ghost Mode” or on-demand reduced-functionality layers that convert smartphones into minimal devices based on predicted risk or geofenced zones. Wearable devices that carry AI assistants can complement screen-free interventions and help keep the individual on track.

\item {\bf Physical DWS zones and living labs}. These can be Signal-limited corridors and urban zones, where participant devices automatically transition into ghost mode. These places enable the study of social interactions, individual attention, physiological markers, and behavioral withdrawal reflexes.

\item {\bf City initiatives} The city could take some initiatives to involve detoxed citizens into social and physical tasks. The following is an indicative list:

\begin{enumerate}
    \item Offline Activity Calendars: Physical boards in public spaces to promote community events
    \item Green Corridors \& Public Fitness areas: Dedicated urban trails and outside gyms
    \item Hobby \& Skill Exchanges: Community meetups in libraries and cultural centers
    \item Community \& Citizen Science: Hands-on projects that enhance local connections
    \item Events in Local Parks: City-sponsored markets, concerts, and festivals without digital distractions
    \item Volunteering Programs: Directing community energy into local supporting roles
\end{enumerate}

\item {\bf National Initiatives}. These include regulatory measures that can be implemented at a national level to support digital well-being.

\end{itemize}

\begin{figure}[!htbp]
    \centering
    \includegraphics[width=1\linewidth]{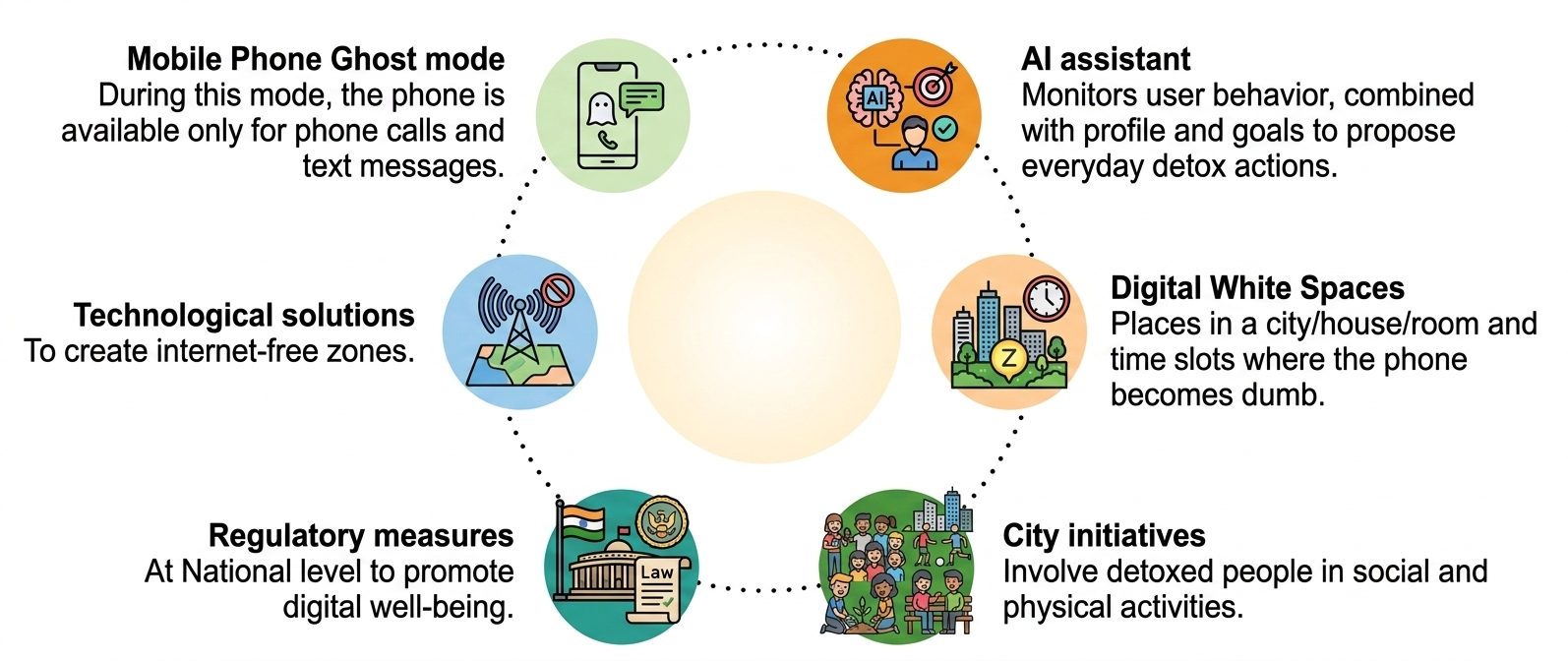}
    \caption{DWS framework}
    \label{fig:figure1}
\end{figure}

All these measures that can form the DWS framework as illustrated in Figure \ref{fig:figure1}

\section{Ethical and Policy considerations}

The proposed DWS concept must ensure GDPR compliance, explicit consent, auditability, and anti-misuse technical safeguards. Any DWS deployment must be reversible, transparent, user-led, and designed to prevent misuse. Some core questions need to be addressed either before or during the implementation of such solutions:

\begin{itemize}
\item  Do short, repeated exposures to DWS reduce daily digital anxiety and improve sustained attention? 

\item  How do behaviorally inferred “points of no return” align with neurophysiological markers? 

\item Which user personas (minors, digital addicts, professionals) benefit or are disadvantaged by DWS? 

\end{itemize}

DWS are not a panacea. They require regulatory, technical, and social support. These include enforceable consent mechanisms, standards for opt-in urban DWS deployments, and interoperable device APIs to enable reversible reduced-functionality modes. Scaling DWS responsibly requires shared standards, rigorous evaluation metrics, and multi-stakeholder governance. The collaboration of engineers, psychologists, and the public administration is crucial for this initiative to be effective.

\begin{figure}[!htbp]
    \centering
    \includegraphics[width=1\linewidth]{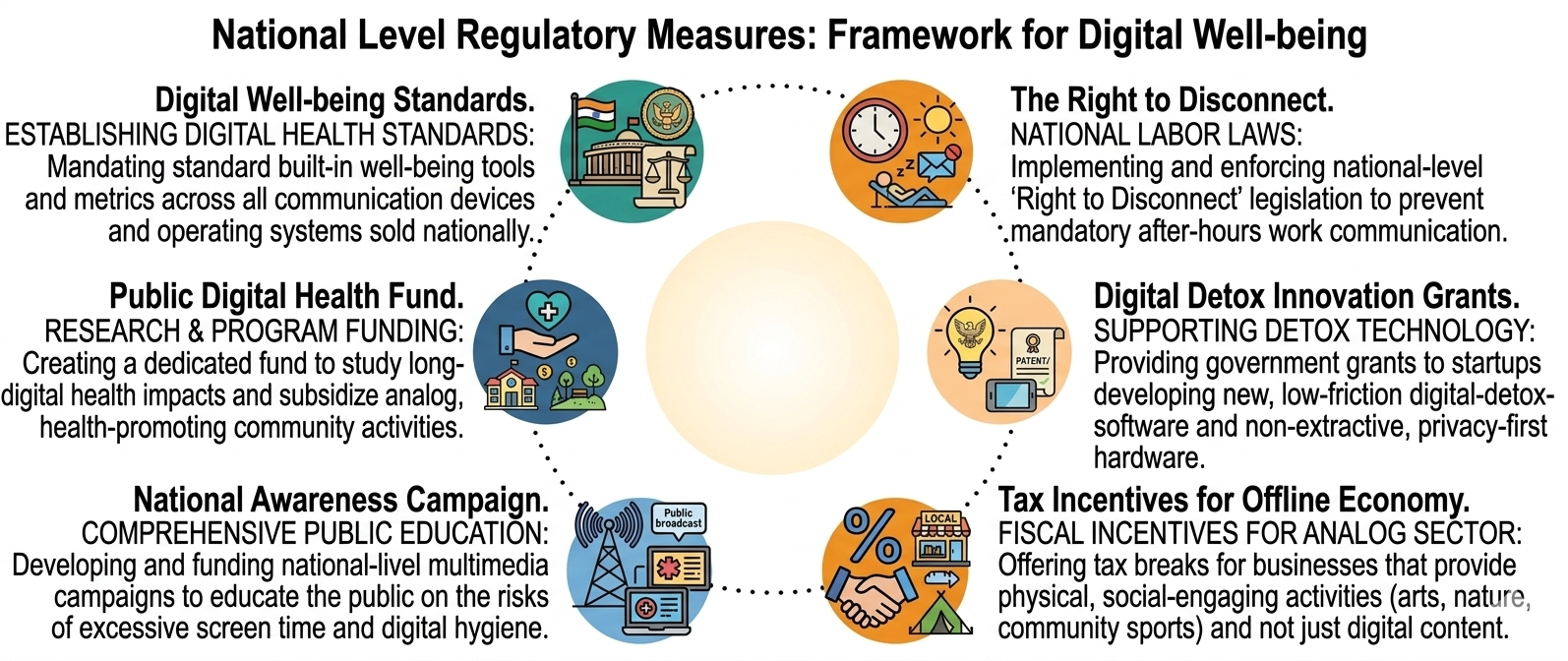}
    \caption{Regulatory measures and initiatives at National level}
    \label{fig:figure2}
\end{figure}

At the national level, each country can take some measures to help the DWS concept. The first step would be to establish clear, unified guidelines for devices and applications, mandating standard built-in well-being tools. This could ensure that every citizen has consistent access to core health features without needing specialized apps. Also, every device could be forced to have the ghost mode as a standard feature. On top of this, implementing clear national labor regulations that define off-hours could be in the right way. This makes it illegal for employers to enforce or expect connectivity (e.g., answering emails) outside of agreed work times. Moreover, developing and funding a major national media campaign (TV, radio, billboards) to educate the public on the long-term impact of screen time and the value of offline activity should follow the regulatory initiatives. Other measures could be the creation of a specific national fund to research digital health and directly subsidize in-person wellness programs and nature conservation,  national-level tax breaks and financial incentives for businesses, and government grants for startups that develop novel, effective digital detox tools and screen-time management solutions (See Figure \ref{fig:figure2}).

\subsection{Regulatory, Legal, and Telecom Feasibility}

Deploying infrastructure-assisted DWS requires navigating strict telecom laws, privacy directives, and emergency accessibility standards:

\begin{itemize}
    \item Telecommunications Regulations \& RF Jamming Prohibitions

{\bf Legal Barrier}: National telecom regulations (e.g., FCC in the US, BEREC in the EU, and national spectrum authorities) strictly prohibit active radio frequency (RF) jamming or cellular signal blocking, as it disrupts public communication channels and violates licensed spectrum usage. 

{\bf Technical Solution}: DWS never utilizes active frequency jamming or network signal blocking. Instead, infrastructure-level nodes function as encrypted, opt-in radio beacons (e.g., Bluetooth Low Energy or geofenced cellular broadcast IDs). When a device with the user-approved DWS app detects the beacon, the app locally activates "Ghost Mode." Unparticipating citizens experience zero network degradation, ensuring full compliance with international telecom laws. 

 \item Privacy Law and Data Autonomy (GDPR / CCPA Compliance)
 
{\bf Legal Barrier}:  Privacy regulations (such as the EU General Data Protection Regulation) prohibit unauthorized tracking of user location, behavioral monitoring, or processing of telemetry without explicit, informed consent. 

{\bf Technical Solution}: DWS relies on a Privacy-by-Design architecture. All behavioral sensing (e.g., scrolling patterns, app session durations) and AI-driven detection of usage loops occur locally on the user's device (Edge AI). Only anonymous, aggregate statistical metrics are shared with municipal platforms. Users retain the right to withdraw consent, toggle exemptions, or disable DWS integration at any moment. 

\item  Public Safety and Emergency Access Requirements 

{\bf Legal Barrier}: Public safety mandates dictate that access to emergency services (e.g., E911, 112) and Advanced Mobile Location (AML) data must never be compromised or delayed by software or hardware locks. 

{\bf Technical Solution}: Safety protocols take strict precedence over DWS modes. Voice lines for emergency calling remain permanently active. If an emergency call is initiated, the DWS application immediately overrides "Ghost Mode," restoring full high-precision GPS, cellular data, and emergency triangulation protocols without user intervention.

\begin{figure}[!htbp]
    \centering
    \includegraphics[width=1\linewidth]{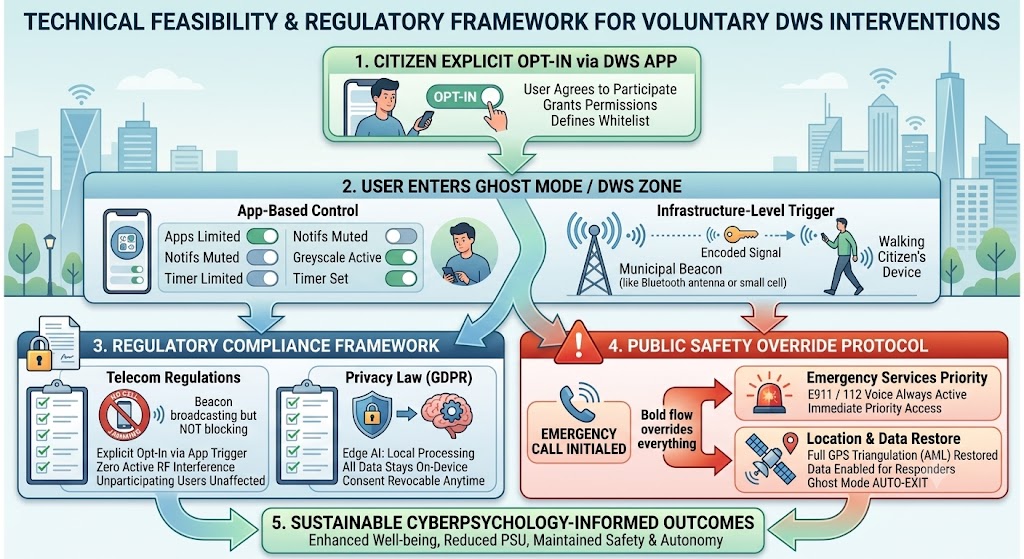}
    \caption {Regulatory, Legal and Technical Issues }
    \label{fig:legal}
\end{figure}
 
\end{itemize}

\section{DWS challenges and Risks}

We tried to identify the main challenges and risks that would arise from the implmentation of the DWS concept and we present them here. For every problem we also proposed a tentative solution, though all these are primivite and need to be further analyzed.

\begin{itemize}
    \item	{\bf Delayed Emergency Response}. Even if the network allows phone calls, losing data services means losing precise location tracking. Emergency services often rely on data-driven GPS triangulation to find people who don't know exactly where they are. A tentative solution for this problem would be the existence of Emergency Override Protocols. City authorities can implement a geofenced "emergency bypass" that temporarily disables the tower's ghost mode in the event of a localized incident.
    
	\item {\bf Navigation Blackouts} Travelers, tourists, or cyclists unfamiliar with the area could get lost instantly if their digital maps, real-time transit updates, or ride-sharing apps suddenly stop working. A tentative solution for this would be the development of a tailored whitelist for every user.   In that concept "Ghost mode" doesn't always have to mean zero data. It can mean curated data. The system can be configured to allow only specific, non-addictive utility apps to function, such as Google Maps, Apple Maps, or local transit apps, while completely blocking social media, email, and browsers, depending on the needs of the user.

    \item {\bf The "Opt-In" problem} If participation is strictly voluntary, it would be sometimes difficult to distinguish between the users who participate in the concept. In case the tower broadcasts a blanket signal to everyone on a DWS route, it could cause some of them to be disconnected. A tentative solution is the use of a dedicated application on the phone in order to participate. To protect user autonomy, citizens who wish to participate will need download a city-approved DWS app. When the tower broadcasts its signal, only devices with the app installed will respond and enter the tailored ghost mode.

    \item {\bf	Cybersecurity Risks} Bad actors could potentially reverse-engineer or spoof the DWS tower's signal to create localized "jamming" zones, cutting off data communication in parts of the city for malicious purposes. Moreover they could try to steal data that are collected from the users in order to perform tailored social engineering  or identity theft attacks. To prevent bad actors from spoofing the tower and jamming signals, the DWS broadcast signal must use encrypted, rotating security keys managed by a central entity and verified by mobile carriers. Moreover, as stated above most of the data of the user will be stored locally to his mobile phone while only aggregated data will be shared with the DWS platform for further analysis. Finally, to mitigate identity attacks during disconnection windows, major identity platforms could integrate APIs that detect a device's DWS status, allowing them to automatically freeze high-risk transactions or enforce time-delay locks if unauthorized actions are attempted from external IP addresses while a user is confirmed to be in ghost mode.
    
\end{itemize}

\begin{figure}[!htbp]
    \centering
    \includegraphics[width=0.9\linewidth]{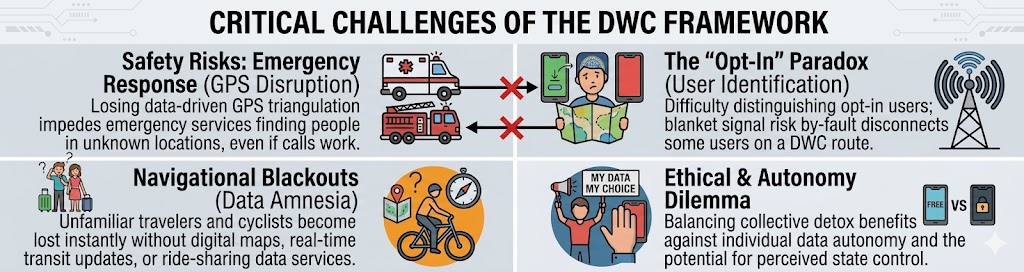}
    \caption{Critical Challenges and Risks of the DWS framework}
    \label{fig:risks}
\end{figure}

In general the DWS concept comes with many new challenges that need to be addressed during design or deployment in order to make it attractive to the citizens. Not all challenges and risks are stated here, neither all tentative solutions that can be used against these risks, since it is a primitive look at this new idea. 

\subsection{Mobile Platform Vulnerabilities and Security Threats}

This subsection presents in detail cybersecurity risks that DWS can raise.  The technical feasibility of deploying socio-technical interventions like Digital White Spaces (DWS) and device-level "Ghost Mode" relies on a trusted mobile platform architecture. However, modern smartphones and wearable devices introduce severe security threats and privacy vulnerabilities across hardware, sensing layers, application ecosystems, and user authentication protocols.

\subsubsection{Authentication and Biometrics Under Hardware Side-Channel Threats}

While on-device "Ghost Mode" activation and privacy-preserving data aggregation depend on robust user authentication, mobile biometric mechanisms remain vulnerable to physical and side-channel exploitation. Beyond traditional software spoofing, physical sensor hardware introduces non-trivial threat vectors. For instance, Ni et al. \cite{ni2023recovering} demonstrated that in-display optical fingerprint sensors radiate electromagnetic signal during the imaging process; attackers can exploit these side channels to reconstruct high-resolution fingerprint images without physical device access or elevated privileges. 

To mitigate biometrics-only vulnerabilities, multi-factor authentication (MFA) architectures are increasingly deployed. Chen et al. (2022) \cite{chen2022swipepass}proposed *SwipePass*, an acoustic-based second-factor authentication mechanism leveraging built-in smartphone speakers and microphones to capture physical touch characteristics. While acoustic sensing offers low-cost, continuous multi-factor verification, acoustic side channels also introduce dual-use risks where ambient audio sampling could be exploited for covert keystroke logging or gesture profiling if platform permissions are mismanaged. Also recently scholras have proposed a dynamic multi-factor authentication (MFA) mechanism based on blockchain technology and honeytokens \cite{papaspirou2025secure}.

\subsubsection{Privacy Leakage Through Unprivileged Sensing and Data Aggregation}
The DWS framework relies on localized, privacy-preserving behavioral monitoring (e.g., session durations, scroll dynamics) to detect "points of no return" in usage loops. However, mobile application ecosystems frequently leak sensitive user telemetry through auxiliary data channels without requiring explicit GPS or high-risk permissions. Meteriz-Yildiran et al. (2022) \cite{meteriz2022learning} revealed that publicly shared profiles in popular fitness applications allow adversaries to infer exact user locations and movement trajectories with high precision by exploiting topological elevation maps. This demonstrates that even high-level, aggregated metadata collected by detox or health-tracking applications can compromise spatial privacy if non-essential telemetry is broadcast.

\subsubsection{Emerging Ecosystems and Dataset Benchmarking}
As digital detox and remote intervention frameworks extend into immersive, augmented reality (AR), virtual reality (VR), and cross-platform mobile environments (e.g., wearable AI assistants), the attack surface expands significantly. Addressing security challenges across emerging mobile platforms requires standardized benchmarking datasets to evaluate threat vectors such as sensor eavesdropping, unauthorized tracking, and resource abuse. Alghamdi et al. (2024) \cite{alghamdi2024xr} introduced *xr-droid*, a benchmark dataset designed to evaluate security, system performance, and threat identification across AR/VR and mobile Android ecosystems. Benchmarking tools like *xr-droid* are essential for validating DWS software components before deployment in real-world living labs or municipal smart city networks.

To ensure long-term trust, DWS infrastructure must enforce strict platform isolation, zero-trust hardware attestation, and local edge processing—preventing raw biometric, acoustic, or spatial telemetry from being exposed to local side-channel attacks or unprivileged third-party applications.

\section{Pilot IMplementation}

While the concept of Digital White Space (DWS) offers a promising approach for reducing user screen time, its effectiveness will be validated through real-world pilot deployments. A structured evaluation framework would enable researchers, policymakers, and technology providers to assess both the technical feasibility and societal impact of the proposed DWS framework.

A  preliminary small pilot deployment could be conducted in controlled environments such as university campuses, public libraries, museums, and followed later by experiments in urban parks, or wellness centers. Participants would voluntarily install the DWS application on their mobile devices after having consented to participate in the study. When entering a designated DWS zone, devices would automatically transition into the Ghost Mode, restricting access to the white list of applications of the user while maintaining essential communication and emergency services.

\begin{figure}[!htbp]
    \centering
    \includegraphics[width=0.9\linewidth]{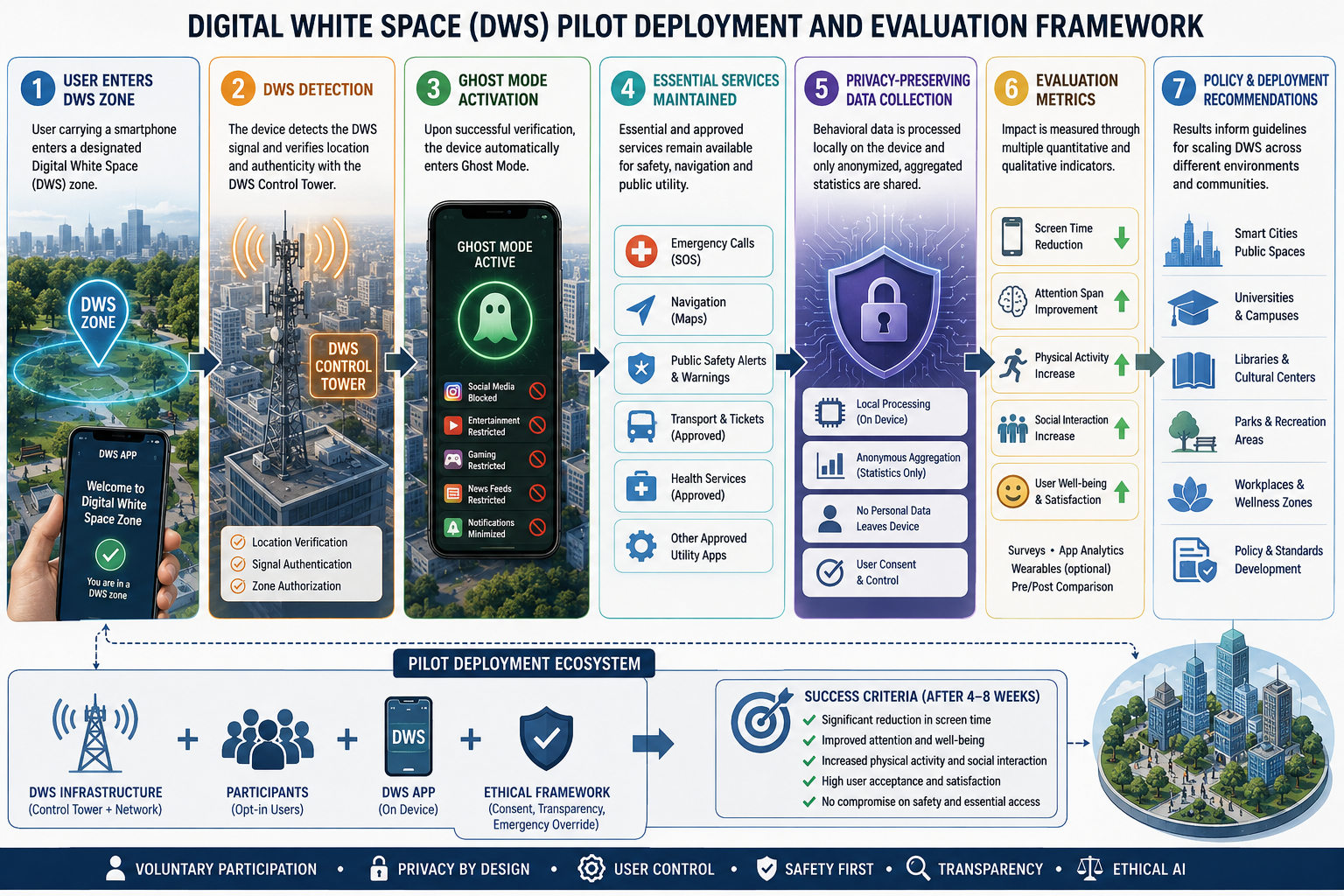}
    \caption{A DWS pilot roadmap}
    \label{fig:roadmap}
\end{figure}

The evaluation process could include both technical and behavioral metrics. Technical metrics may include correct authentication, location detection accuracy, battery consumption, and network performance. Behavioral metrics may include mean and total reduction in screen time, attention span, social engagement, physical activity levels, and other self-reported well-being indicators. Surveys conducted before and after participation could provide valuable insights into user acceptance, perceived usefulness, and overall satisfaction.

A successful pilot deployment would provide evidence regarding the practicality of DWS implementations and help identify optimal deployment strategies for future large-scale smart city applications. The findings could also contribute to the development of regulatory frameworks and technical standards governing Digital White Space ecosystems.

\subsection{Target Demographics and Application Environments}
DWSs cater to diverse user segments experiencing distinct technological pressures.

\begin{table}[htbp]
\centering
\caption{Target Demographics, Technological Stressors, and Core Needs Addressed by Ghost Mode Zones}
\label{tab:ghost_mode_target_groups}
\small
\begin{tabularx}{\textwidth}{>{\raggedright\arraybackslash}p{3.2cm} >{\raggedright\arraybackslash}X >{\raggedright\arraybackslash}X}
\toprule
\textbf{Target Demographic} & \textbf{Dominant Technological Stressors} & \textbf{Core Need Addressed by Ghost Mode Zones} \\
\midrule
\textbf{Youth} \newline (Ages 15--25) & High social media reliance, continuous online presence, elevated FoMO. & Enhancing self-regulation, focus, and authentic offline relationships. \\
\addlinespace
\textbf{Young Professionals} \newline (Ages 25--45) & Hyper-connectivity, off-hours work communication, telework burnout. & Mental decompression, stress reduction, and work-life boundaries. \\
\addlinespace
\textbf{Families} & Parallel screen use, reduced quality shared time. & Co-created offline play, storytelling, and screen-free social bonding. \\
\addlinespace
\textbf{High-Cognitive-Load Workers} & Continuous task-switching (executives, researchers, educators, healthcare professionals). & Cognitive restoration, focus recovery, and mental clarity. \\
\bottomrule
\end{tabularx}
\end{table}

These interventions can be integrated into existing urban infrastructure:
\begin{itemize}
    \item Urban Parks \& Greenery: National gardens, neighborhood parks, coastal walkways, and botanical gardens
    \item Institutional \& Work Spaces: Public libraries, cultural centers, wellness hubs, and designated corporate quiet zones
    \item Pilot "Islands": Micro-installations or pop-up offline hubs deployed in high-density urban areas
\end{itemize}

DWS utilize a flexible, three-tiered operational model:

\begin{enumerate}[label={Level \arabic*}]
    \item Passive Environmental Nudging: Soft environmental design characterized by no public Wi-Fi, ambient physical signage, "conversation benches", and screen-free seating areas
        \item Active Voluntary Disconnection: Facilitated offline experiences where users manually trigger device Ghost Mode, select a defined disconnection period, and engage in provided offline media or activities
    \item Scheduled Ghost Mode Events: Organized community initiatives, such as silent walks, digital detox days, guided mindfulness sessions, family offline days, and interactive storytelling
\end{enumerate}

\section{conclusions}
Digital White Spaces combine cyberpsychology with engineering and policy to offer a structured, ethical approach for mitigating mobile-phone addiction and restoring physical activities. Moving forward, coordinated experiments, guided by strong ethical safeguards and inclusive stakeholder engagement, can test whether intermittent, context-aware disconnection can become a scalable public-health tool.

\bibliographystyle{unsrt}
\bibliography{biblio}

\end{document}